\documentclass[journal,12pt,onecolumn,draftclsnofoot,]{IEEEtran}
\usepackage{graphicx}
\usepackage{float}
\usepackage{amsmath}
\usepackage{cases}
\usepackage{amsfonts}
\usepackage{algorithm}
\usepackage{color}
\usepackage{algorithmic}
\usepackage{cite}
\begin{document}

\title{Wireless Transmission of Big Data: Data-Oriented Performance Limits and Their Applications}

\author{
Hong-Chuan Yang, \textit{Senior Member, IEEE} and Mohamed-Slim Alouini, \textit{Fellow, IEEE}
\thanks{This work was supported in part by NSERC Discovery Grant}
\thanks{H.-C. Yang is with the Department of Electrical and Computer Engineering, University of Victoria, Victoria, BC V8W 2Y2, Canada (e-mail: hy@uvic,ca).}
\thanks{M.-S. Alouini is with the
Computer, Electrical, and Mathematical Sciences and Engineering (CEMSE) Division, King Abdullah University of Science
and Technology (KAUST), Thuwal 23955, Saudi Arabia (e-mail: slim.alouini@kaust.edu.sa).}
}


\IEEEtitleabstractindextext{
\begin{abstract}
The growing popularity of big data and Internet of Things (IoT) applications bring new challenges to the wireless communication community. Wireless transmission systems should more efficiently support the large amount of data traffics from diverse types of information sources. In this article, we introduce a novel data-oriented approach for the design and optimization of wireless transmission strategies. Specifically, we define new performance metrics for individual data transmission session and apply them to compare two popular channel-adaptive transmission strategies. We develop several interesting and somewhat counterintuitive observations on these transmission strategies, which would not be possible with conventional approach. We also present several interesting future research directions that are worth pursuing with the data-oriented approach. 
\end{abstract}

}

\maketitle

\IEEEdisplaynontitleabstractindextext
\IEEEpeerreviewmaketitle

\section{Introduction}

\IEEEPARstart{W}{e} are in an era of big data. Data are generated and collected at an accelerating rate. The timely processing, delivery, and analysis of these data will bring huge social and economic benefit \cite{7492186,7809016}. With the intensive ongoing deployment of wireless communication systems, most big data will be transmitted over the air. In fact, smart mobile devices contribute significantly to the generation of big data. 
The ever-growing  Internet of Things (IoT) devices serve as another source of big data for wireless transmission. The supporting of big data transmission presents several technical challenges to wireless system design, including spectrum efficiency enhancement of radio  access network (RAN), capacity provision of fronthaul/backhaul links, and network architecture improvement for traffic scalability. To effectively support various big data and IoT applications, future wireless systems need to optimize their transmission strategies for a large amount of data from diverse sources. 

There have been significant development in digital wireless transmission technologies over the past two decades. Various advanced transmission technologies, including multiple antenna (MIMO/massive MIMO) transmission\cite{foschini98, larsson14}, channel adaptive transmission\cite{chua98}, cooperative relay transmission\cite{sendonaris1, laneman2}, cognitive radio transmission\cite{788210}, and extreme bandwidth transmission (e.g. millimeter-wave, terahertz , and optical wireless transmission), are developed and deployed to meet the growing demand for high data rate wireless services. These transmission technologies were typically designed with the goal of enhancing or approaching the capacity limits of wireless channels, usually characterized by ergodic capacity and outage capacity. Ergodic capacity specifies the upper limits of average transmission rate over fading channels, whereas outage capacity corresponds to the largest instantaneous transmission rate of the channel under a specific outage probability constraint. The rationale is that enhancing the channel quality will necessarily improve the quality of service experienced by individual transmission session on average. Such \emph{channel oriented} approach worked quite effectively so far and has successfully facilitate the delivery of high-quality wireless services.

Current wireless systems typically apply the same transmission strategy to all transmission sessions over the wireless channel channel. With the application of advanced transmission technologies, the properties of the channel, e.g. average data rate and average error rate, will be improved, which usually translates to better average quality of service experienced by individual sessions. Such channel oriented design works perfectly well for transmission sessions with long duration, such as phone calls and video streaming. Meanwhile, the channel oriented design ignores the specifics of individual transmission sessions, such as the traffic characteristics and the prevailing network/channel condition. When the transmission sessions are short, the quality of service experienced by individual sessions vary dramatically around the average. In particular, it was recently shown that the transmission time of a fixed amount of data with adaptive transmission over fading channels vary considerably  around its average \cite{wangyang18}. With the growing popularity of IoT devices and big data applications, future wireless systems need to support increasing number of short transmission sessions, initiated for example by sensor nodes.  


To further improve the efficiency of wireless transmission systems, especially for IoT and big data applications, we need to study wireless transmission technologies from a new perspective. In this article, we advocate the perspective of individual transmission sessions.  Intuitively, we expect that the performance/efficiency of wireless transmission can be further enhanced if the transmission strategy is optimized for each transmission session based on the traffic characteristics and operating environment. 
Motivated by this intuition, we propose a novel \emph{data-oriented} approach for wireless transmission system design. Specifically, when a certain amount of data is available for transmission, we will decide the transmission strategy in an optimal fashion. For example, should power adaptation should be applied together with rate adaptation or not? Should cooperative relaying be activated or not? What multiple antenna transmission structure should apply? The transmission strategy will be adjusted for each data transmission session according to the traffic characteristics and the channel/network conditions. The rationale for the data oriented approach is that optimizing the transmission strategy of individual sessions will directly improve the quality of service for them and will in turn enhance the transmission efficiency of the overall system. We believe the proposed data-oriented approach will facilitate the design of efficient transmission solutions for big data and IoT applications. 

There are many challenges to be addressed for the new data-oriented approach. We first need to define suitable metrics to quantify the quality of service experienced by individual data transmission session. We also need to establish the performance limits from the data transmission perspective and use them as guideline to optimize the transmission strategy. In this article, we present some initial investigation of the data-oriented approach. In particular, we introduce two new data-orient performance limits for individual data transmission sessions. As an initial application of these performance limits, we compare the performance of two popular channel adaptive transmission strategies over fading channels when the channel state information is available at the transmitter. Finally, we discuss several promising application and future research directions for the data-oriented approach.

\section{Data oriented performance limits}

Ergodic capacity and outage capacity are well-known performance limits for wireless transmission over fading channels. Ergodic capacity applies to the scenario that the transmission will experience all possible fading states. It characterizes the largest possible transmission rate that the channel can support over fast fading environment or extremely long transmission duration. Outage capacity, on the other hand, is applicable to slow fading environment and specifies the largest transmission rate that the channel can support under a specific outage probability requirement. In general, outage capacity specifies the instantaneous capacity limit, i.e. within a channel coherence time, over which the channel realization is highly correlated, whereas ergodic capacity dictates the average rate limit over a long duration, e.g. orders of magnitude larger than a coherence time. These \emph{channel-oriented} performance limits can not fully describe the quality of service experienced by individual data transmission session, especially for big data and IoT applications. 

The wireless transmission of big data often involves multiple channel coherence time. Consider, for example, the indoor transmission of an AR/VR video over IEEE 802.11ac WiFi. The typical file size of AR/VR videos is around 4 Gbits whereas the peak download speed of 802.11ac can reach 2.5 Gbps. As such, the video transmission can finish in 1.6 second on average. The channel coherence time of typical operating environment for WiFi is around 200 ms. Therefore, the transmission will last for about eight coherence time periods. As another example, consider the outdoor transmission of high-quality image over an LTE link. The file size of the image can be several hundred of Kbits after compression and the transmission speed of LTE link can reach up to Mbps.  The transmission will last for about hundreds of milliseconds, which entails several coherence time periods for a typical coherence time value of tens of milliseconds for outdoor environment. 

Ergodic capacity can only characterize the quality of service experienced by a particular wireless transmission session in an average sense. The actual transmission service experienced by the data transmission session depends heavily on the prevailing channel realization. The effective transmission rate of particular session will vary dramatically around the average rate. To more effectively characterize the quality of transmission service, we raise the following questions: Given a certain amount of data to be transmitted, what is the chance that it will be successfully transmitted within a fixed time duration? Given the available temporal-spectral resource, what is the largest amount of data that can be transmitted over the channel reliably? Outage capacity characterizes instantaneous rate limit and is only applicable for transmission sessions that last less than one channel coherence time. Outage capacity can not be generalized to transmission spanning multiple channel coherence time. To effective address the above design questions, we need suitable new  \emph{data-oriented} performance metrics. In the following, we present two new performance limits. 



\subsection{Minimum transmission time}

The fundamental service requirement of many big data and IoT applications is to transmit a certain amount of data to its destination in a timely fashion. As such, we define a data-oriented metric, minimum transmission time (MTT), as the minimum time duration required to transmit a certain amount of data over wireless channels. Let $H$ denote the amount of data to be transmitted. The MTT will be a function of $H$, denoted by $T_\mathrm{min}(H)$. For a given $H$ value, MTT will vary with the channel bandwidth,  the channel realization, and the adopted transmission strategy. 
When $H$ is relatively small and the data transmission completes in one channel coherence time, MTT $T_\mathrm{min}(H)$ depends on the instantaneous channel realization. With optimal rate adaption (ORA) \cite{ADD}, the maximum transmission rate over a channel coherence time is equal to $B\cdot \log_2(1+\gamma)$, where $B$ is the channel bandwidth and $\gamma$ is the instantaneous received SNR. MTT can then be calculated as $H/B\log_2(1+\gamma)$, which will vary with the received SNR $\gamma$. When, on the other hand, $H$ is very large and the data transmission involves many coherence time, MTT can be calculated using the ergodic capacity of the channel, given by $\overline{C} = \int_0^\infty B\log_2(1+\gamma) p_\gamma(\gamma) d\gamma$, as $\mathrm{MTT} = H/\overline{C}$, which is a constant value. 

To address the earlier design questions, we define the delay outage rate (DOR) as the probability that MTT for a certain amount of data is greater than a threshold duration. In particular, DOR is mathematically given by 
 $\mathrm{DOR} = \Pr[T_\mathrm{min}(H) > T_\mathrm{th}]$, where $T_\mathrm{th}$ denotes the threshold duration. In informational theoretical sense, $H$ represents the amount of information contained in the data. $T_\mathrm{th}$ can be related to the delay requirement of the data to be transmitted. As such, DOR serves as an statistical measure for the quality of service experienced by individual data transmission session. For example, the DOR for data transmission within a channel coherence time with ORA can be calculated as
\begin{equation}
\mathrm{DOR}^\mathrm{ora}
= \Pr\left[\gamma<\exp(\frac{H\ln(2)}{BT_\mathrm{th}})-1\right],
\end{equation}
which specifies the performance lower limit for the transmission time without power adaptation. 

When the data transmission lasts more than one channel coherence time, as is the case for big data transmission, DOR analysis becomes more challenging. Assuming ORA over block fading, where the received SNR remains constant over each coherence time of $T_c$ and changes to an independent value afterwards, MTT is less than $L \cdot T_c$ if $\sum_{l=1}^L T_c B\log_2(1+\gamma_l) > H$, where $\gamma_l$ is the received SNR over the $l$th $T_c$. 
As such, the DOR for the case of $T_\mathrm{th}=LT_c$ can be calculated as 
 \begin{equation}
\mathrm{DOR}^\mathrm{ora}
= \Pr\left[\sum_{l=1}^L T_c B\log_2(1+\gamma_l) > H\right].
\end{equation}
To accurately evaluate the above probability, we need the statistical distribution of the sum of $L$ independent random variables $T_c B\log_2(1+\gamma_l)$, which may be solvable using the Fox $H$ function \cite{5478750}. The DOR analysis for general scenarios would be an interesting research problem for further investigation. 

\subsection{Maximum entropy throughput}

Wireless communication systems accommodate the service requirements of big data and IoT applications by allocating certain spectral-temporal resource. The characterization of the amount of data that can be successfully transmitted over a certain spectral-temporal resource block would be instrumental to the design of resource allocation algorithms. As such, we define maximum entropy throughput (MET) as the maximum amount of information that can be transmitted over a certain time duration and channel bandwidth. Mathematically, we denote MET  by $H_\mathrm{max}(T, B)$, which is a function of the time duration $T$ and the channel bandwidth $B$. Here, $T$ represents an arbitrary time duration, with value ranging from less than one coherence time $T_c$ to many $T_c$'s. For given $T$ and $B$ values, MET will depends on the channel realization and the adopted transmission strategy. For example, when $T$ is larger than $T_c$ by order of magnitude, MET can be calculated as $H_\mathrm{max}(T, B)=T\overline{C}$, where $\overline{C}$ is the ergodic capacity of the channel. On the other hand, if $T$ is smaller than $T_c$ and ORA is applied, then MET can be calculated using the instantaneous channel capacity as $H_\mathrm{max}(T, B)=TB\log_2(1+\gamma)$. The analysis of MET for the case that $T$ spans multiple $T_c$ will be more involved. 

Since MET is generally varying with the channel realization, we define the information outage rate (IOR)  as the probability that MET over a certain time duration is less than a threshold entropy value, denoted by $H_\mathrm{th}$. Mathematically, IOR is given by $\Pr[H_\mathrm{max}(T, B) < H_\mathrm{th}]$. 
The IOR analysis will be instrumental to the design of efficient resource allocation algorithms. For data traffic with stringent delay requirement, it is desirable to allocate sufficient temporal-spectral resource such that the data transmission can successfully complete within delay constraint with high probability. Apparently, the IOR analysis requires the statistics of  $H_\mathrm{max}(T, B)$, which depends on the channel statistics and the adopted transmission strategy. For example, when $T$ is very large compared with $T_c$, IOR will be equal to 0 if 
the channel ergodic capacity $\overline{C}$ is greater than $H_\mathrm{th}/T$. Meanwhile, when $T$ is less than $T_c$,  the IOR can be calculated, assuming that the system applies ORA, as
 \begin{equation}
\mathrm{IOR}^\mathrm{ora}
= \Pr[\gamma<\exp({\frac{\ln(2)H_\mathrm{th}}{BT}})-1].
\end{equation}
For the scenario that $T$ involves multiple $T_c$'s, the IOR will be equal to the probability that MET is less than $H_\mathrm{th}$, the evaluation of which will requires the distribution of the sum of $N$ independent random variables.  Further investigation of IOR will be an interesting topic for future research.

\section{Transmission Strategy Comparison with CSIT}

As an application of data-oriented performance limits introduced in previous section, we now compare ORA and optimal power and rate adaptation (OPRA) strategies over a point-to-point wireless channel.  When the full channel state information is available at the transmitter (CSIT), wireless transmission with ORA can achieve the ergodic capacity of fading channels. It has also been established that OPRA transmission can further enhance the capacity of fading wireless channel with water filling  power allocation\cite{ADD}. In particular, the resulting OPRA capacity is considerably higher than the ergodic capacity over low SNR region\cite{6186721}.  Will OPRA transmission still outperform ORA transmission from the perspective of a particular data transmission session? We can apply data-oriented performance metrics introduced in previous section, i.e. DOR and IOR, to answer this question. 

The DOR and IOR with ORA over slow fading environment are given in Eq. (1) and Eq. (3), respectively. With OPRA, the instantaneous channel capacity becomes $B\log_2(\gamma/\gamma_T)$ when the received SNR is great than a threshold SNR $\gamma_T$ and zero otherwise. The threshold SNR $\gamma_T$ is determined to satisfy the average transmit power constraint with the optimal water-filling power allocation policy. As such, the DOR with OPRA for slow fading environment is calculated as 
\begin{equation}
\mathrm{DOR}^\mathrm{opra}
= \Pr[\gamma<\gamma_T \exp({\frac{H\ln(2)}{BT_\mathrm{th}}})],
\end{equation}
and the IOR for OPRA transmission over slow fading channel is determined as
\begin{equation}
\mathrm{IOR}^\mathrm{opra}
= \Pr[\gamma<\gamma_T \exp({\frac{H_\mathrm{th}\ln(2)}{BT}})]. 
\end{equation}

\begin{figure}[t]
\begin{center}
\includegraphics[height=9.2cm]{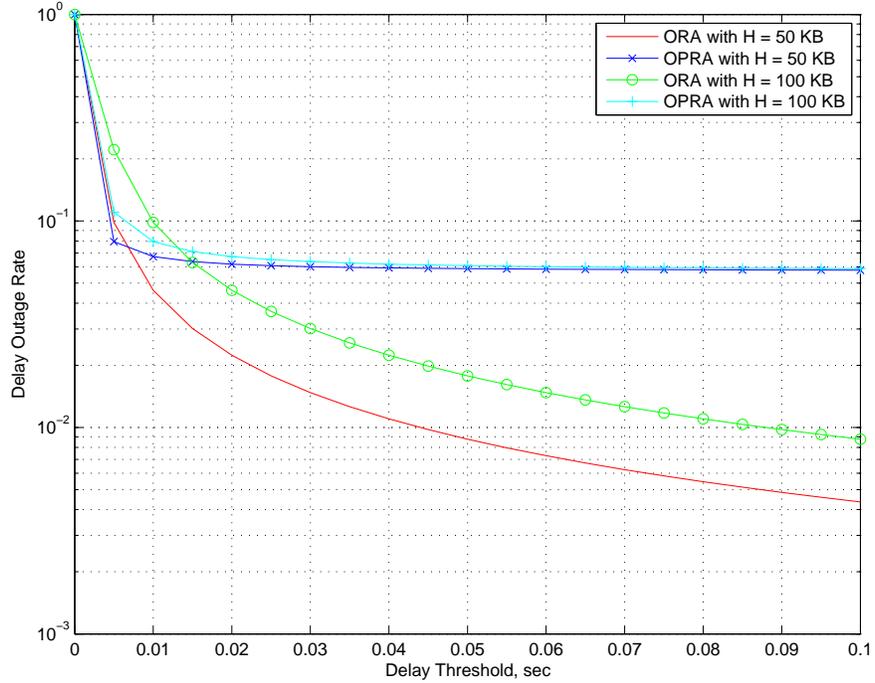}
\end{center}
\caption{Delay outage rate of ORA and OPRA transmission over slow Rayleigh fading channel ($B =$ 20 MHz, and $\overline{\gamma} =$ 6 dB).}
\label{DORHdiff}
\end{figure}

\begin{figure}[t]
\begin{center}
\includegraphics[height=9.2cm]{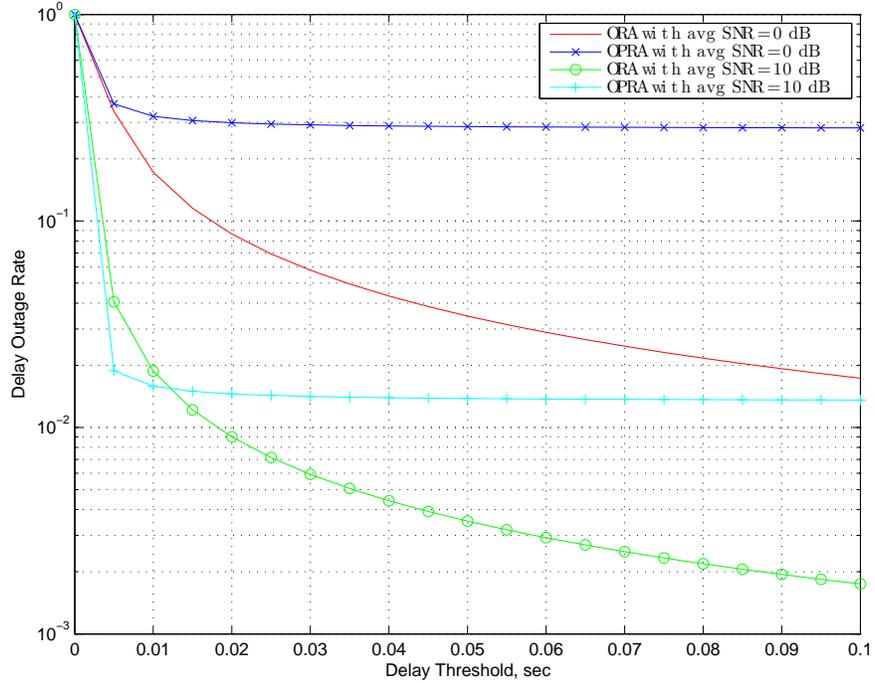}
\end{center}
\caption{Effect of average SNR on the delay outage rate of ORA and OPRA transmission over slow Rayleigh fading channel ($B =$ 20 MHz, and $H$ = 50 KB).}
\label{DORgbardiff}
\end{figure}

Fig. \ref{DORHdiff} compares the DOR performance of ORA and OPRA transmission strategies over slow Rayleigh fading channel. In particular, we plot DOR of both strategies as function of the delay threshold $T_\mathrm{th}$ for different data amount $H$. We can see that for both choices of $H$ values, there is a mixed behavior between the DOR performance of ORA and OPRA.  Specifically, when the delay threshold is small, OPRA leads to smaller DOR than ORA. When the threshold duration becomes larger, the DOR with ORA transmission improves and becomes much smaller than that with OPRA. In fact, the DOR of OPRA converges to a fixed value when delay threshold becomes very large, which is equal to the probability of no transmission with OPRA. Fig. \ref{DORgbardiff} illustrates the effect of the average received SNR on the DOR performance. We can see that when the average SNR is small, ORA always achieve smaller DOR than OPRA, which holds the transmission with higher probability. When the average SNR increases, the DOR performance of OPRA improves, but still is worse than that of ORA when the delay threshold is large. Note that from the conventional ergodic capacity perspective, OPRA considerably outperform ORA over low SNR regime. We observe from the DOR comparison, however, that OPRA is not always the better strategy from the perspective of individual transmission session. OPRA is preferred over ORA when the delay requirement is very stringent or the channel quality is favorable.

\begin{figure}[t]
\begin{center}
\includegraphics[height=9.2cm]{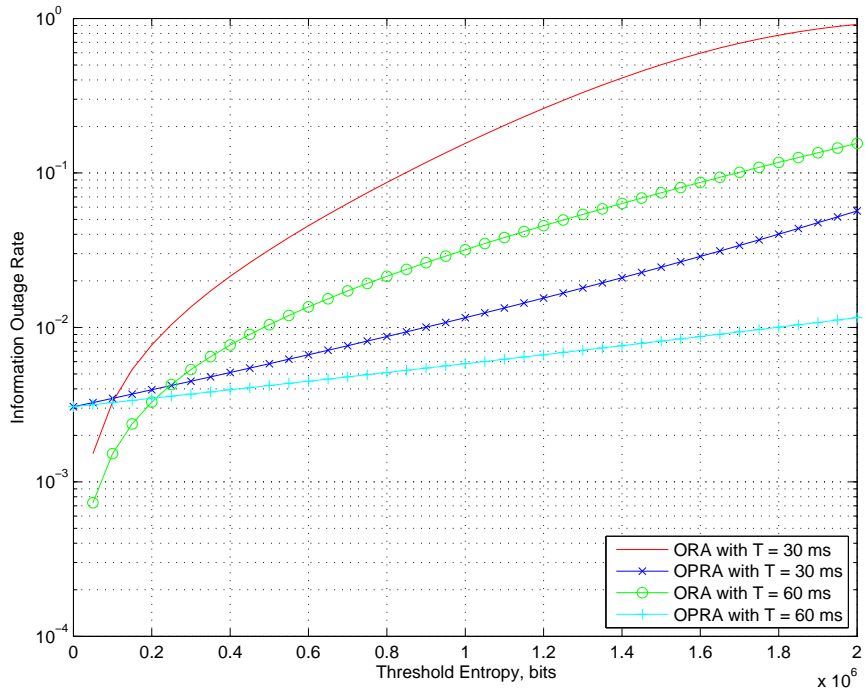}
\end{center}
\caption{Information outage rate of ORA and OPRA transmission over slow Rayleigh fading channel ($B =$ 20 MHz, and $\overline{\gamma} =$ 6 dB).}
\label{IORTdiff}
\end{figure}

\begin{figure}[t]
\begin{center}
\includegraphics[height=9.2cm]{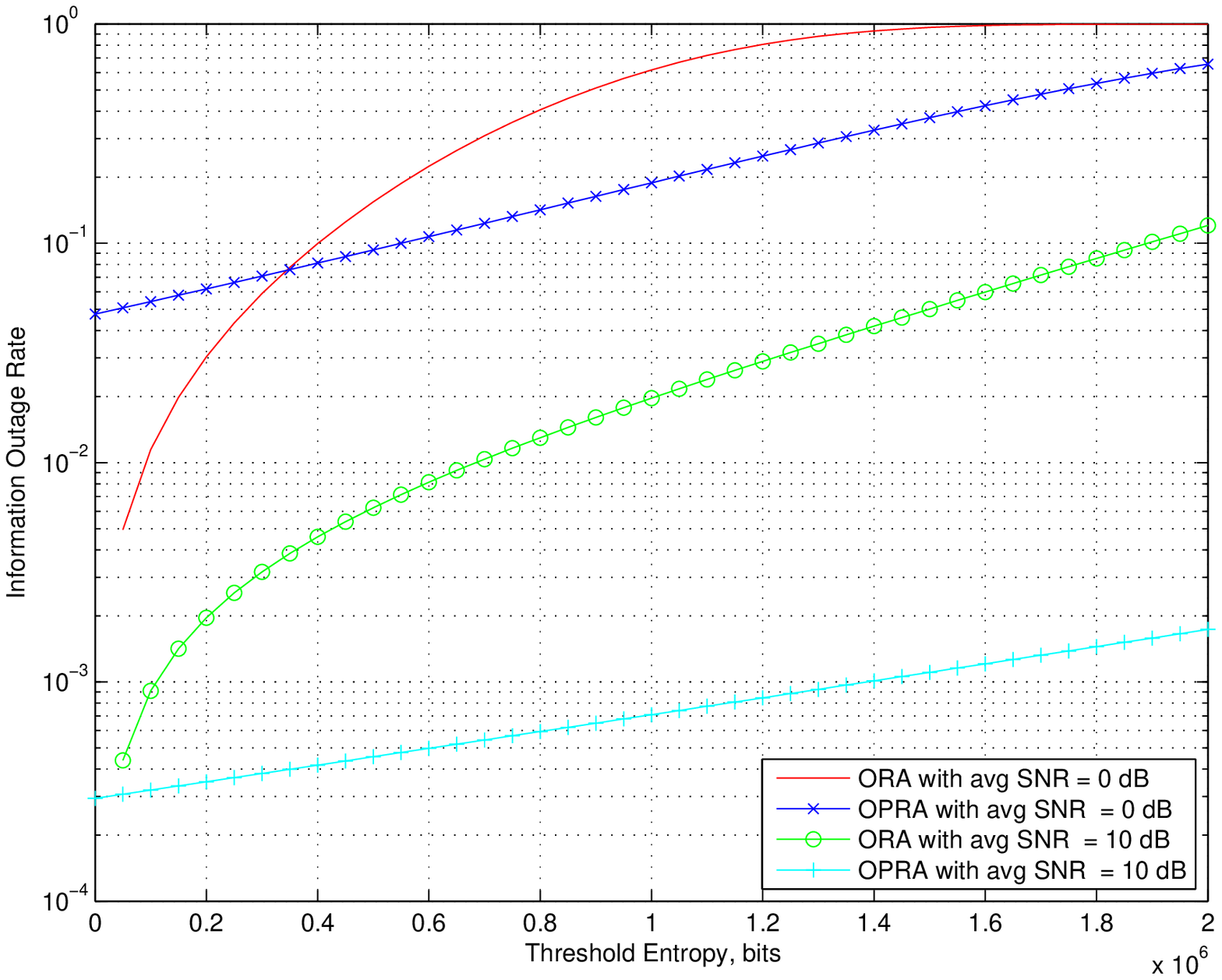}
\end{center}
\caption{Effect of average SNR on the information outage rate of ORA and OPRA transmission over slow Rayleigh fading channel ($B =$ 20 MHz, and $T =$ 30 ms).}
\label{IORgbardiff}
\end{figure}

We now compare the IOR performance of ORA and OPRA transmission strategies over slow Rayleigh fading channel.  In Fig. \ref{IORTdiff}, we plot IOR of both strategies as function of the entropy threshold $H_\mathrm{th}$ for different time duration $T$. We again observe a mixed behavior between the IOR performance of ORA and OPRA.  Specifically, when the entropy threshold is small, ORA leads to smaller IOR value than OPRA. When the entropy threshold becomes larger, the IOR with ORA transmission increases and quickly becomes larger than that with OPRA. In fact, the IOR of OPRA steadily increases from  a fixed value, which is equal to the probability of no transmission, for both time duration values. Fig. \ref{IORgbardiff} illustrates the effect of the average received SNR on the IOR performance. We can see that when the average SNR is small, ORA outperforms OPRA in terms of IOR for a wider range of entropy threshold value. When the average SNR increases, the IOR performance of OPRA improves significantly and is better than that of ORA unless $H_\mathrm{th}$ is extremely small. We can conclude from this comparison that from the perspective of individual transmission session, OPRA transmission strategy should be used over high SNR scenario and/or when a large amount of data is to be transmitted.

\section{Further applications}

The new data oriented performance limits characterize the performance of individual data transmission sessions over fading wireless channels. In particular, MTT prescribes the smallest transmission delay possible when transmitting a certain amount of data over fading channels, where as MET signifies the largest amount information that can be transmitted over a temporal-spectral resource block. Given the time-varying nature of wireless fading channel, these performance limits are described in a statistical sense, in terms of DOR and IOR, respectively. By specifying the best possible performance for individual transmission session, these performance limits will find many important applications for the design and optimization of wireless transmission strategies. 

In previous section, we compared two channel adaptive transmission strategies for the slow fading scenario with CSIT using the data oriented metrics. The design insights developed therein can readily apply to the transmission scheme optimization for IoT traffics, which are typically brief and sporadic. On the other hand, big data applications tend to generate large volume of data traffics, the transmission of which may last multiple channel coherence time. The data oriented analysis for big data traffic will be a challenging but rewarding future research topic. An initial investigation on the transmission time of big data with discrete rate adaptation over fading channels has been recently reported \cite{wangyang18}. 

The data oriented approach can also apply to the design and optimization of practical transmission strategies with limited CSIT.  Adaptive modulation and coding (AMC)  and automatic repeat request (ARQ) are two popular transmission strategies that explores limited feedback from the receiver. AMC adapts the transmission rate for a certain reliability requirement whereas ARQ enhance the reliability with retransmission. The joint design of AMC and ARQ has been investigated in the literature\cite{7604086}. With the proposed data-oriented approach, we can study these two transmission strategies and their joint design from a brand new perspective. Such study will create new design insights and leads to novel transmission strategy for the limited CSIT scenario. 

The MTT analysis characterizes the lower limit for the transmission time of practical data transmission.  For point-to-point links, the transmission time is inverse proportional to the service rate of the transmission system. The queuing delay performance for wireless transmission can be analyzed using the first-order and second-order statistics of transmission time \cite{7111368}. Our data-oriented characterization can apply to develop the upper bound of the queuing performance over point-to-point link. Meanwhile, transmission time is directly related to the channel occupancy of each transmission session. The statistical characterization of the transmission can be used to optimize random access protocols. 

The growing popularity of big data and IoT applications will create an unprecedented amount of traffic with diverse service requirement. The wireless system need to apply efficient resource allocation algorithms to accommodate such new demands in an effective manner. The MET characterization will provide valuable guidelines to the design and optimization of resource allocation algorithms. For example, 3GPP adopted the scheduled uplink approach for IoT provisions, which involves a random access stage for scheduling request\cite{IoT3GPP17}. Only terminals succeeded in this stage will be allocated with resource block. Therefore, it is critical to allocate sufficient resource blocks for each terminals such that their transmission can finish with high probability. With the data-oriented approach, we can enhance the performance of such resource allocation algorithms. 


\section{Concluding Remarks}

In this article, we present a novel data-oriented approach for wireless transmission system design and analysis. We target at the transmission strategy design and optimization for individual data transmission session, according to the traffic characteristic and operating condition. In particular, we introduce two data-oriented performance limits to characterize arbitrary wireless data transmission. As their initial application, we compare well-known channel adaptive transmission strategies for CSIT scenario, namely ORA and OPRA. We observe that while OPRA always outperform ORA from the ergodic capacity perspective, OPRA is not always the preferred transmission strategy from the individual data transmission session perspective. ORA can have a better chance to deliver the data to the destination over slow fading channel when the average channel quality is poor. As such, the data-oriented approach can bring interesting new insights to wireless transmission over fading channels. 

This article serves as an initial introduction to the data-oriented approach for wireless transmission strategy design. There are many important aspects to be addressed. The limited and no CSI at transmitter scenarios are of practical interest. The data oriented characterization of big data transmission requires further investigation. Finally, these characterization can be readily applied to the queuing analysis and resource allocation for wireless systems. We expect that the data-oriented design will greatly facilitate the design of efficient wireless transmission strategies for big data and IoT applications. 

\bibliographystyle{IEEEtran}
\bibliography{IEEEabrv,IEEETED}

\end{document}